\documentclass[a4paper, 10pt]{article}
\usepackage[backend=biber,style=ieee,citestyle=ieee]{biblatex}
\usepackage{tikz}
\usepackage{url}
\usepackage{cancel}
\usepackage{blindtext}
\usepackage{graphicx}
\usepackage{pgfplots}
\usepackage{hyperref}
\usepackage{color,colortbl}
\usepackage{xspace}
\usepackage{amssymb}
\usepackage[cmex10]{amsmath}
\usepackage{mathtools}
\mathtoolsset{showonlyrefs,showmanualtags}  
\usepackage{xcolor}
\usepackage[normalem]{ulem}
\usepackage[footnotesize]{caption}
\usepackage{subcaption}
\usepackage{amsthm}

\newcommand{\vs}{\vspace{-7.5pt}}
\newcommand{\hvs}{\vspace{-3pt}}
\newcommand{\htex}{2in}
\newcommand{\sizep}{1pt}
\newcommand{\bs}{\boldsymbol}
\newcommand{\bb}{\mathbb}

\newcommand{\ts}{\textstyle}

\newcommand{\ie}{\emph{i.e.}, }
\newcommand{\eg}{\emph{e.g.}, }

\newcommand{\mS}{\mathcal{S}}
\newcommand{\iid}{%
    \ifmmode% math mode
        \mathrm{i.i.d.}%
    \else%
        i.i.d.\@\xspace%
    \fi%
}

\DeclareMathOperator{\im}{\mathsf{i}}
\DeclareMathOperator{\supp}{supp}

\usetikzlibrary{through,calc}

\definecolor{bblue}{HTML}{4F81BD}
\definecolor{rred}{HTML}{C0504D}
\definecolor{ggreen}{HTML}{9BBB59}
\definecolor{ppurple}{HTML}{9F4C7C}

\definecolor{mDarkBrown}{HTML}{604c38}
\definecolor{mDarkTeal}{HTML}{23373b}
\definecolor{mMediumTeal}{HTML}{205A65}

\definecolor{mLightTeal}{HTML}{41a3c3}

\definecolor{mLightBrown}{HTML}{EB811B}
\definecolor{mMediumBrown}{HTML}{C87A2F}

\theoremstyle{definition}

\usepackage[colorinlistoftodos]{todonotes}

\definecolor{TF}{HTML}{008080}

\usepackage[top=1.5cm, left=1.5cm, right=1.5cm, bottom=1.5cm]{geometry}

\title{\vspace{-1.75cm}\textit{One Bit to Rule Them All:}\\ Binarizing the Reconstruction in 1-bit Compressive Sensing}

\author{\small Thomas Feuillen, Mike Davies, Luc Vandendorpe, and Laurent Jacques\\
\small UCLouvain, Louvain-la-Neuve, Belgium\\ \small Edinburgh University, United Kingdom\vs \vs}
\date{\empty} % no need for a date

\renewenvironment{abstract}{\bf\small {\em\ Abstract---}}{}
\definecolor{redn}{rgb}{0.83921568627451,0.152941176470588,0.156862745098039}
\definecolor{bluen}{rgb}{0.12156862745098,0.466666666666667,0.705882352941177}
\newcommand{\greenn}{green!50.19607843137255!black}
\begin{document}

\maketitle

\begin{abstract} 
This work focuses on the reconstruction of sparse signals from their 1-bit measurements. The context is the one of 1-bit compressive sensing where the measurements amount to quantizing (dithered) random projections. Our main contribution shows that, in addition to the measurement process, we can additionally reconstruct the signal with a binarization of the sensing matrix. This binary representation of both the measurements and sensing matrix can dramatically simplify the hardware architecture on embedded systems, enabling cheaper and more power efficient alternatives. Within this framework, given a sensing matrix respecting the restricted isometry property (RIP), we prove that for any sparse signal the quantized projected back-projection (QPBP) algorithm achieves a reconstruction error decaying like $O(m^{-\frac{1}{2}})$ when the number of measurements $m$ increases. Simulations highlight the practicality of the developed scheme for different sensing scenarios, including random partial Fourier sensing.
\end{abstract}

\section{Introduction}

Since its advent in 2004 \cite{4472240, donoho}, Compressive Sensing (CS) has focused on reducing the data rate needed for low-complexity signals (\eg sparse) while still allowing their estimation.
%One of the main focus of Compressive Sensing  (CS)\cite{candes06} has  been on reducing the requirements needed to perform reconstructions of signals from measurements.
 By introducing new properties on the measurement process such as the RIP \cite{intro2CS,Baraniuk2008ASP}, it was first shown that the number of measurements necessary can be related to the information rate of the signal (\eg its sparsity).
 Further works focused on reducing the resolution of the acquired data, creating more cost efficient platform with lower resolution Analog to Digital Converter (ADCs) or reducing the storage needed on the processing platform \cite{jlpea8020012}.
 In \cite{xu2018quantized}, the authors proposed  adding a random dither and then quantizing the compressive measurements. By leveraging this dither they showed that reconstruction error of PBP fed with Quantized measurements (PBPQ) decreases like $O(m^{-\frac{1}{2}})$ when the number of measurements $m$ increases.  

In this paper we aim at further reducing the requirements on the hardware by performing the PBP reconstruction with a 1-bit quantification of the measurement matrix, what we refer as Quantized Backprojection on 1-bit quantized measurement (QPBPQ). Apart from the obvious storage gain provided by the binary representation of the matrix, the Backprojection process itself can be made more efficient. For example, scalar multiplications between 1-bit variables become simple XOR logic gates, which can be efficiently computed or even implemented on embedded platforms. This has the potential to yield extremely efficient architectures \cite{complexhadamard}. A similar simplified processing was studied in \cite{7079881} however, they provided no proof of a reconstruction error that decreases with the number of measurements $m$.

This work is also partly connected to the study of mixed operator in CS \cite{herman2010mixed}, and on sensing matrix corruption \cite{Herman2010GeneralDA}. In \cite{herman2010mixed} the authors considered performing the reconstruction using an operator deviating from the original one, in this context the mismatch between the two operators is seen as multiplicative noise. The main difference with our work (apart from the use of highly quantized measurements) is that, in \cite{herman2010mixed}, no assumption is made on the structure of the mismatch which results in quite stringent conditions for the reconstruction of signals. The developed theory in \cite{herman2010mixed} only allows for $\approx 5\%$ (in a $\ell_2$ sense) discrepancy between the two operators. Whereas our proposed scheme creates a mismatch that is as big (in amplitude) as the maximum component of the matrix, but leverages the added dither to obtain reconstruction error that scales down with an increasing number of measurements.

The claims of this paper are the following: (\textit{i}) we prove a uniform bound on the degradation between PBPQ and QPBPQ, and show that it scales as  $O(m^{-\frac{1}{2}})$, provided $\bs \Phi $ follows the RIP;
(\textit{ii}) we validate the results through Monte-Carlo simulations for Fourier and Gaussian complex matrices, which also highlight the necessity of adding the dither before the quantization;
 (\textit{iii}) we show empirically that this scheme could be extended to matrices that have a factorized representation (such as Fast Fourier Transform \cite{Magoarou2016FlexibleMS}), enabling for further gain in the implementation.

\section{System Model \& Reconstruction}

In CS, we consider the sensing model\hvs 
\begin{equation}
\label{eq:lin-mod}
\ts \bs y = \bs \Phi \bs x,\hvs
\end{equation}
where $\bs y \in \mathbb{C}^m$ is the complex measurement vector, and $\bs x \in \mathbb{C}^N$ is a complex sparse vector. We assume $\bs \Phi \in \mathbb{C}^{m \times N}$ to respect the RIP$(s,\delta)$ for a sparsity $s$ and a constant $0 <\delta < 1$: for all $s$-sparse vectors $\bs x \in \mathbb{C}^N$,\hvs
\begin{equation}
\ts (1-\delta)\| \bs x \|^2_2 \leq \frac{1}{m}\| \bs \Phi \bs x \|^2_2 \leq (1+\delta) \| \bs x \|^2_2.\hvs
\end{equation}
This property has been proven for random (sub) Gaussian matrices \cite{Baraniuk2008ASP}, random partial Fourier matrices, and other structured sensing matrices \cite{intro2CS}.

In this work, we alter \eqref{eq:lin-mod} with a dithered uniform quantizer $\mathcal{Q}_\alpha(\cdot)$ of resolution $\alpha >0$ \cite{xu2018quantized}. For $\bs u \in \bb R^m$, this quantizer, applied componentwise (entrywise) on vectors (matrices), reads $\mathcal{Q}_\alpha(\bs u) := \lfloor \frac{\bs u + \bs \xi}{\alpha } \rfloor \alpha +\frac{\alpha}{2}$. The dither $\bs \xi \in \mathbb{R}^m$ follows a uniform distribution over $[-\frac{\alpha }{2},\frac{\alpha }{2}]^m$. For complex vectors, we quantize the real and imaginary parts separately with independent dithers \cite{Feuillen20181bitLS,Feuillen2018QuantityOQ}. If $\|\bs u\|_\infty \leq \alpha/2$, the quantization is binary in the real and the imaginary components, \ie $\mathcal{Q}_\alpha(\bs u) \in \frac{\alpha}{2} \{\pm 1 \pm \im\}^m$, and $\|\mathcal{Q}_\alpha (\bs u)\|_\infty\leq {\alpha}/{ \sqrt 2}$. 

In this context, we focus of the projected back-projection algorithm \cite{intro2CS} for the signal reconstruction, namely the signal estimate\hvs 
\begin{equation}
\ts \hat{\bs x}= \frac{1}{m} \mathsf{H}_s (\bs A^T   \bs b),\hvs
\end{equation}
where $\mathsf{H}_s(\bs u)$ is the hard thresholding operator that zeroes all but the $s$ biggest components of $\bs u$ in amplitude.
Depending on the combination of $\bs A$ and $\bs b$, we obtain different schemes. In classical PBP, $\bs A = \bs \Phi$ and $\bs b =\bs  \Phi \bs  x$ \cite{intro2CS}, and in PBPQ \cite{xu2018quantized} $\bs A = \bs \Phi$ and $\bs b = \mathcal{Q} (\bs \Phi \bs  x)$. Here, we quantize both $\bs y$ and $\bs \Phi$ (with resolutions $\epsilon$ and $\nu$, respectively), and set $\bs A = \mathcal{Q}_\nu(\bs \Phi)$ and $\bs b = \mathcal{Q}_\epsilon(\bs \Phi\bs x)$, what we refer as the QPBPQ algorithm.

\section{Bound on Quantized PBP}

The $\ell_2$ reconstruction error of QPBPQ is given by 
\begin{equation}
\ts
\|\bs x -\hat{\bs x} \|_2=\|\bs x -\frac{1}{m}\left( \mathcal{Q}_\nu(\bs \Phi) ^T \mathcal{Q}_\epsilon(\bs \Phi \bs x)\right)_\mathcal{S} \|_2
\end{equation}
where $\epsilon \ge 2\|\bs \Phi \bs x\|_\infty$ (and $\nu \ge 2\|\bs \Phi\|_\infty$) is the quantization resolution such that the elements of $\bs \Phi \bs x$ (resp. $\bs \Phi$) can be represented with 1-bit and $\mathcal{S}:=\supp(\mathsf{H}_s[\mathcal{Q}_\nu(\bs{ \Phi}^T)\mathcal{Q}_\epsilon(\bs \Phi \bs x)])$. For brevity, we define hereafter $\bs z := \mathcal{Q}_\epsilon(\bs \Phi \bs x)$, $\bs \Psi := \mathcal{Q}_\nu(\bs \Phi)$.

We show that, when $m$ is large enough and for a matrix $\bs \Phi$ respecting the RIP$(2s,\delta)$, the reconstruction error of any unit $s$-sparse vector $\bs x$ is bounded as $\|\bs x -\hat{\bs x} \|_2 = O(\sqrt{ s/ m})$. Below, the values $C,c > 0$ are universal constants that may change from one instance to another. 

For brevity's sake we only provide a proof sketch. The full proof is inspired by \cite{xu2018quantized, Baraniuk2008ASP}. We first note that\vs 
\begin{align}
\ts \|\bs x -\hat{\bs x} \|_2 & \leq \|\bs x-\frac{1}{m} \big(\bs \Psi^T  \bs z\big)_\mathcal{S} \|_2 
\leq 2\|\bs x- \frac{1}{m}\big(\bs \Psi^T \bs z\big)_\mathcal{T} \|_2\\
\ts & \leq 2\|\bs x - \frac{1}{m}\big(\bs \Phi^T  \bs z\big)_\mathcal{T} \|_2 + 2\|\frac{1}{m}\big((\bs \Phi -\bs \Psi)^T \bs z\big)_\mathcal{T}\|_2,\vs \label{eq:deg}
\end{align}
with $\mathcal{T}:= \mS \cup \supp{(\bs x)}$, $|\mathcal{T}|\leq 2s$. 

In \eqref{eq:deg}, the first term, which amounts to the PBPQ reconstruction over the extended support of QPBPQ, is upper-bounded by $\mathcal{O}(\delta(1+\epsilon )m^{-\frac{1}{2}})$ \cite[Cor.~3.9]{xu2018quantized}. For the second term, we must show that, for $\bs y = \mathcal{Q}_\epsilon(\bs \Phi \bs u)$, the quantity $\|\frac{1}{m}\big((\bs \Phi -\bs \Psi)^T \bs y \big)_{\mathcal{T}'}\|_2$ is small for all unit $s$-sparse vectors $\bs u$ and any support $\mathcal{T}'$, with $|\mathcal{T}'|\leq 2s$. Let us first consider $\bs u$, and thus $\bs y$, fixed. From the properties of the dithering \cite{xu2018quantized}, $\bb E [\bs \Psi_i^T \bs y] =\sum_j^m \mathbb{E}(\Psi_{ij}^*) \bs y = \bs  \Phi_{i}^T  \bs y$. Given $i \in [N]$ and $\gamma>0$, Hoeffding's inequality provides
\begin{equation}
\ts
\bb P\big(|\bs \Psi_i^T \bs y - \bs \Phi_i^T \bs y |>m\gamma \big)\leq C \exp\big(\frac{-c m \gamma^2 }{  \epsilon^2 \nu^2  }\big),
\end{equation}
since $\|\bs \Psi_i - \bs \Phi_i\|_\infty \leq \nu$. Therefore, by union bound, we can extend this inequality to any $i \in [N]$, so that
\begin{equation}
\ts \| ((\bs \Psi - \bs \Phi)^T  \bs y)_{\mathcal{T}'}\|_2^2  = \sum_{i \in \mathcal{T}'} |\bs \Psi_i^T \bs y - \bs \Phi_i^T \bs y|^2\leq 2 m^2 s \gamma^2, \vs
\label{eq:nonuniform}
\end{equation}
with probability of failure $p_\gamma\leq  C n \exp(\frac{- c m \gamma^2 }{ \epsilon^2 \nu^2  })$. 

To extend \eqref{eq:nonuniform} to all bounded $s$-sparse vectors $\bs u$, we define a $\rho$-covering of bounded size $|\mathcal{J}_{\rho}|\leq (\frac{e N}{s})^{s} (1+\frac{2}{\rho})^{2s}$ of the set of $s$-sparse vectors \cite{intro2CS}. By union bound, \eqref{eq:nonuniform} holds for all $\bs u \in \mathcal{J}_\rho$ with probability exceeding $1 - C n \exp(\log |\mathcal{J}_{\rho}| - \frac{c m \gamma^2 }{ \epsilon^2 \nu^2})$. The quantizer prevents us to extend by continuity this last property to all bounded $s$-sparse signals. However, for any bounded $s$-sparse vector $\bs v$ whose closest point in $\mathcal{J}_\rho$ is $\bs u$, we can leverage Lemma 6.1 in \cite{xu2018quantized} to bound the number of non-zero components in the vector $\mathcal{Q}_\epsilon(\bs \Phi \bs v) - \mathcal{Q}_\epsilon(\bs \Phi \bs u)$ to a small fraction of $m$. If $\rho$ is proportional to $\gamma / (\epsilon \nu^2 \sqrt m)$, this fraction is bounded by $c \gamma/\epsilon^2\nu^2$. 

We thus use that value of $\rho$ and gather all the events above (with a probability of failure upper-bounded by the sum of the previous failure probabilities). Finally, provided that\hvs 
$$
\ts m \geq c \gamma^{-2}\,s \log (\frac{n \sqrt m \epsilon \nu^2}{\gamma}),\hvs
$$
we have, for all unit $s$-sparse vectors $\bs v$,
\begin{equation}
\ts \|\frac{1}{m}(\bs \Psi - \bs \Phi)^T_{\mathcal{T}'} \mathcal{Q}(\bs \Phi \bs v)  \|_2 \leq C \sqrt{s} \gamma(1 + \frac{1}{\epsilon\nu})\vs
\end{equation}
with probability exceeding $1 - C\exp(-c\frac{\gamma^2 m}{ \epsilon^2 \nu^2})$. Placing this bound in \eqref{eq:deg} and saturating the requirement on $m$ provides the announced decay for the reconstruction error of QPBPQ. 

\section{Simulation}

We assess the quality of the developed scheme by performing Monte-Carlo simulations. We carried out 100 runs for different numbers of measurements and sparsity levels.

\begin{figure}[!h]
\centering% This file was created by matplotlib2tikz v0.7.5.
\begin{tikzpicture}

\begin{axis}[
width=0.95\columnwidth,
height=0.8*\htex,
tick align=outside,
tick pos=left,
x grid style={white!69.01960784313725!black},
xlabel={\scriptsize $\log_2(\frac{m}{N})$},
xmin=-4., xmax=7.,
xtick style={color=black},
y grid style={white!69.01960784313725!black},
ylabel={\scriptsize $10\log_{10}(\|\bs x_0 -\frac{\hat{\bs x}}{\|\hat{\bs x}\|}\|_2)$},
ymin=-15, ymax=2.,
ytick style={color=black}
]
\addplot [very thick, redn, dotted, mark=*, mark size=\sizep, mark options={solid}]
table {%
-6 0.241474095478519
-5 -1.10145055319168
-4 -2.90765480715534
-3 -5.59076322475684
-2 -8.21420701218262
-1 -11.867043283144
-0.415037499278844 -14.734343834088
0 -128.354667380193
1 -128.35450087596
2 -128.354595420946
3 -128.353391426738
4 -128.35562451095
7 -128.33555734773
};
%\addlegendentry{PBP 2}
\addplot [very thick, redn, dashed, mark=*, mark size=\sizep, mark options={solid}]
table {%
-6 1.45474035079104
-5 1.15683135306468
-4 0.571271629867275
-3 -0.839954486239839
-2 -3.10014552137859
-1 -5.45107304201199
-0.415037499278844 -7.10371923354453
0 -8.26114871983039
1 -9.96988913748727
2 -12.1762470807959
3 -13.746557871004
4 -15.4491154214827
7 -19.8497481151098
};
%\addlegendentry{PBPQ 2}
\addplot [very thick, redn, mark=*, mark size=\sizep, mark options={solid}]
table {%
-6 1.49856962184164
-5 1.46534933479945
-4 1.15423566494016
-3 0.371280095339912
-2 -0.884788796247854
-1 -2.606727610066
-0.415037499278844 -4.5313940523761
0 -5.64066752059623
1 -7.63488846205958
2 -9.88652107406233
3 -11.6025440356639
4 -13.3097001845682
7 -17.6666469196101
};
%\addlegendentry{QPBPQ 2}
%\addplot [very thick, red, mark=triangle*, mark size=\sizep, mark options={solid,rotate=180}]
%table {%
%-6 1.43766452168618
%-5 1.30023566943378
%-4 0.949259025329172
%-3 -0.473524504319625
%-2 -2.33239396032183
%-1 -4.36124408987264
%-0.415037499278844 -5.21093041503267
%0 -5.74053280377122
%1 -6.38196991060428
%2 -6.57897385908438
%3 -6.80815945744627
%4 -7.06983595244749
%7 -7.32476863871192
%};
%\addlegendentry{NQPBPQ 2}
\addplot [very thick, bluen, dotted, mark=*, mark size=\sizep, mark options={solid}]
table {%
-6 1.33381093050858
-5 1.0923803323427
-4 0.398884498708512
-3 -0.742318997313749
-2 -2.52618671483376
-1 -5.4421128382334
-0.415037499278844 -8.3234784714475
0 -120.441769150585
1 -120.441749843455
2 -120.441740737087
3 -120.4418573225
4 -120.441926276705
7 -120.439944253413
};
%\addlegendentry{PBP 10}
\addplot [very thick, bluen, dashed, mark=*, mark size=\sizep, mark options={solid}]
table {%
-6 1.432529553058
-5 1.41507441661748
-4 1.36763553884576
-3 1.25032928238636
-2 1.0005257872891
-1 0.571046815307976
-0.415037499278844 0.031037570832991
0 -0.256918438730242
1 -1.76147191262009
2 -3.3865270605049
3 -5.20167048234423
4 -7.05706869819744
7 -12.3985141723076
};
%\addlegendentry{PBPQ 10}
\addplot [very thick, bluen, mark=*, mark size=\sizep, mark options={solid}]
table {%
-6 1.45914534157871
-5 1.44770027381255
-4 1.42351702121301
-3 1.38312356155407
-2 1.25740217250255
-1 1.02674582899576
-0.415037499278844 0.839859675217843
0 0.669646219057128
1 -0.309138775133084
2 -1.54685346328904
3 -3.26747509293803
4 -4.98858636005296
7 -10.5385122356583
};
%\addlegendentry{QPBPQ 10}
%\addplot [very thick, blue, mark=triangle*, mark size=\sizep, mark options={solid,rotate=180}]
%table {%
%-6 1.44809422029237
%-5 1.42117156286904
%-4 1.37769906959145
%-3 1.28866317255323
%-2 1.13769283964176
%-1 0.752843193484566
%-0.415037499278844 0.382503802228707
%0 0.0678652708789178
%1 -1.25600015725181
%2 -2.56045420876049
%3 -3.62149361934823
%4 -4.38391313637603
%7 -5.06102898230186
%};
%\addlegendentry{NQPBPQ 10}
\addplot [very thick, white!50.19607843137255!black, dashed, mark=triangle*, mark size=\sizep, mark options={solid,rotate=180}]
table {%
-6 14.9897000433602
-5 13.4845500650403
-4 11.9794000867204
-3 10.4742501084005
-2 8.96910013008056
-1 7.46395015176066
-0.415037499278844 6.58349385648225
0 5.95880017344075
1 4.45365019512085
2 2.94850021680094
3 1.44335023848103
4 -0.0617997398388717
7 -4.57724967479859
};
%\addlegendentry{m**0.5}
\end{axis}

\end{tikzpicture}
\vs
\caption{$\|\bs x- \frac{\hat{\bs x}}{\|\hat{\bs x}\|_2}\|_2$ in dB, for different numbers of measurements ($\log_2(\frac{m}{N})$), the dotted curves are the classic PBP; the dashed, PBPQ and the solid, QPBPQ. The colours represent the sparsity, $s=2$ for red and $s=10$ for blue. The dashed grey line represent the decrease rate of $\mathcal{O}(m^{-\frac{1}{2}})$.\vs}
\label{fig:L2}
\end{figure}
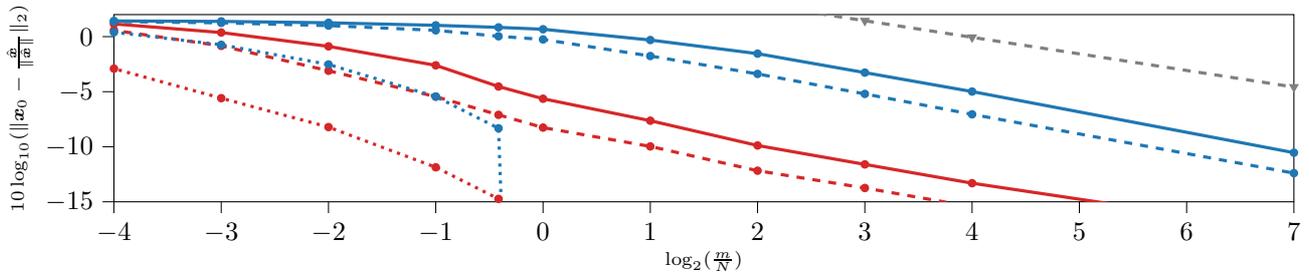

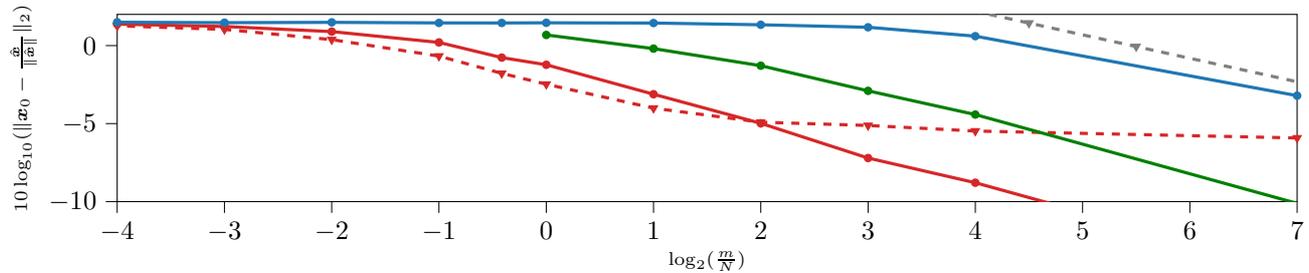
\begin{figure}[!h]
\centering% This file was created by matplotlib2tikz v0.7.5.
\begin{tikzpicture}

\begin{axis}[
width=0.95\columnwidth,
height=0.8*\htex,
tick align=outside,
tick pos=left,
x grid style={white!69.01960784313725!black},
xlabel={\scriptsize $\log_2(\frac{m}{N})$},
xmin=-4., xmax=7.,
xtick style={color=black},
y grid style={white!69.01960784313725!black},
ylabel={\scriptsize $10\log_{10}(\|\bs x_0 -\frac{\hat{\bs x}}{\|\hat{\bs x}\|}\|_2)$},
ymin=-10, ymax=2.,
ytick style={color=black}
]
\addplot [very thick, redn, mark=*, mark size=\sizep, mark options={solid}]
table {%
-6 1.48517686976097
-5 1.43892660000816
-4 1.37564353274378
-3 1.22423594095408
-2 0.896666211528423
-1 0.20267529634947
-0.415037499278844 -0.769833773200269
0 -1.2288796986021
1 -3.11765957548517
2 -4.98273853029524
3 -7.20870431139305
4 -8.79064847656366
7 -14.4935290555085
};
%\addlegendentry{F QPBPQ 4}
\addplot [very thick, redn, dashed, mark=triangle*, mark size=\sizep, mark options={solid,rotate=180}]
table {%
-6 1.45198588571818
-5 1.42984216271507
-4 1.26681169629749
-3 1.0373601690758
-2 0.382333494577598
-1 -0.673978316720579
-0.415037499278844 -1.77964160908325
0 -2.475899669145
1 -4.01144441426059
2 -4.91031559582169
3 -5.11990711041446
4 -5.47777168844128
7 -5.92211989827306
};
%\addlegendentry{F NQPBPQ 4}
\addplot [very thick, \greenn, mark=*, mark size=\sizep, mark options={solid}]
table {%
0 0.681706470579956
1 -0.193674778797465
2 -1.28880871059387
3 -2.90090029448826
4 -4.42154972767383
7 -10.1159813966838
};
%\addlegendentry{fft QPBP 4}
\addplot [very thick, bluen, mark=*, mark size=\sizep, mark options={solid}]
table {%
-6 1.51404474423183
-5 1.49198729783272
-4 1.49738324379245
-3 1.46852489288769
-2 1.48949657874183
-1 1.45428776753018
-0.415037499278844 1.45116058620909
0 1.46038276695985
1 1.44586933252394
2 1.33431409342869
3 1.17470568518486
4 0.60406820720491
7 -3.20987781205989
};
%\addlegendentry{G QPBPQ 4}
\addplot [very thick, white!50.19607843137255!black, dashed, mark=triangle*, mark size=\sizep, mark options={solid,rotate=180}]
table {%
1.5 5.95880017344075
2.5 4.45365019512085
3.5 2.94850021680094
4.5 1.44335023848103
5.5 -0.0617997398388717
8.5 -4.57724967479859
};
%\addlegendentry{m**0.5}
\end{axis}

\end{tikzpicture}

\caption{$\|\bs x- \frac{\hat{\bs x}}{\|\hat{\bs x}\|_2}\|_2$ in dB, for different number of measurement ($\log_2(\frac{m}{N})$), for different schemes with a sparsity of $s=4$, namely QPBPQ with dithering in solid red for Fourier matrix, QPBPQ without dithering in $\bs \Psi$ in dashed red for Fourier matrix, the Quantized FFT in green and the complex Gaussian matrix in blue.\vs}
\label{fig:gfft}
\end{figure}

In Fig. \ref{fig:L2}, we compare the proposed QPBPQ scheme in terms of $\ell_2$-reconstruction error against PBP and PBPQ for two sparsity level, $s=2$ and $s=10$, with a measurement matrix corresponding to a randomly (sub/over)-sampled Fourier transform. As indicated by the developed proof, QPBPQ does scale as $\mathcal{O}(m^{-\frac{1}{2 }})$ and only suffers from a constant loss in dB compared to PBPQ. Interestingly, it seems that the loss of resolution in the measured signal has more impact on the performances than lowering the resolution of the back-projection. 
In Fig. \ref{fig:gfft} we compare different schemes with the same sparsity level of $s=4$. Similarly to what was shown in \cite{xu2018quantized, Feuillen20181bitLS,Feuillen2018QuantityOQ}, we see the dither plays in capital role in the obtained performances. Indeed, while the dithered scheme continues to scale down when $m$ increases, the scheme with the deterministic backprojection seems to slowly saturates. We also see that using complex Gaussian matrices yields similar result, albeit requiring a larger number of measurements to reach the same performances,  compared to Fourier transforms. Finally we compare the 1-bit Backprojection using a quantized Fourier transform with the 1-bit version of the Butterfly Fast Fourier Transform.
% In the presented framework, it corresponds to applying the quantization to each element of a factorized matrix, which represents the different weights of the Fast Fourier transform.
 This corresponds to quantizing independently the $\mathcal{O}( \log{(N)})$ matrices composing the Butterfly factorization  of the Fourier matrix, before sub/over-sampling its rows \cite{Magoarou2016FlexibleMS}. 
  Although it is not yet fully supported by the developed theory, it is interesting to see that this scheme can benefit from relative low losses while still enjoying fast computation thanks to the butterfly algorithm reducing the complexity from $\mathcal{O}(mN)$ to $\mathcal{O}( m \log{N})$.

\section{Conclusion}

\label{sec:conclusion}
We showed that leveraging the effect of a dither can allow the reconstruction of sparse signal using a simplified 1-bit backprojection. The proposed scheme shows a relative low recovery losses compared to other high resolution methods such as PBPQ. 
This fully quantized scheme was shown to have a bounded reconstruction error which decreases with an increasing number of measurements. 
%This was demonstrated using simulations highlighting the practicality of QPBPQ. 
Future work will focus on extending the scheme to noisy settings and fast algorithm such has the Fast Fourier transform and other factorizable matrices. %This proof will also be extended to the reconstruction of high resolution signal using the 1-bit Backprojection (QPBP) in a noisy setting in a future manuscript soon to be published.
 More generally, our approach could be open to the quantization of an iterative reconstruction algorithm, similarly to \cite{8682059}.
\printbibliography
\end{document}